# Interakt—A Multimodal Multisensory Interactive Cognitive Assessment Tool

**Daniel Sonntag**
German Research Center for Artificial Intelligence
Stuhlsatzenhausweg 3
66123 Saarbruecken, Germany

## Abstract

Cognitive assistance may be valuable in applications for doctors and therapists that reduce costs and improve quality in healthcare systems. Use cases and scenarios include the assessment of dementia. In this paper, we present our approach to the (semi-)automatic assessment of dementia.

## Motivation

Neurocognitive testing is to assess the performance of mental capabilities, including for example, memory and attention. Most cognitive assessments used in medicine today are paper-pencil based. These tests are both expensive and time consuming. (A doctor, physiotherapist or psychologist conducts the assessments.) In addition, the results can be biased.

Through the use of digital pen technologies and mobile devices, multimodal multisensory data (e.g., speech and handwriting) could be collected and evaluated. We plan to use time-stamped stroke data from a digital pen. In this paper, we describe a new means of doing objective neurocognitive testing with digital pen technology on normal paper and on tablets, and we address the issue of what role automation could play in designing adaptive multimodal-multisensor interfaces (Oviatt et al. 2017) to support precise medical assessments; an online test environment will be developed, too, where a doctor, physiotherapist, or psychologist is assisted by an automatic assessment tool which classifies time-stamped stroke data from a digital pen in real-time.

We can identify one special target group of multimodal-multisensory interfaces as public sector applications: cognitive assistance in terms of automatically interpreted dementia tests. Personalised assessment systems through gamification and serious games have great potential and can improve quality care in healthcare systems. Our new project Interakt with clincial partners from Charité in Berlin complements previous fundamental research projects for reality orientation and validation dialogue (Sonntag 2015); the design, development and evaluation of interactive technologies aimed at changing users' attitudes or behaviours through persuasion (Sonntag 2016); and clinical data intelligence (Sonntag et al. 2016). Previous approaches of inferring cognitive status from subtle behaviour have been made in a clock drawing test, a simple pencil and paper test that has proven useful in helping to diagnose cognitive dysfunction such as Alzheimer's disease. The test is a de facto standard in clinical practice as a screening tool to differentiate normal individuals from those with cognitive impairment and has been digitised in a first version with a digital pen only recently (Davis et al. 2014; Souillard-Mandar et al. 2016). As pointed out in (Davis et al. 2014), the use of a tablet and stylus may also distort results by its different ergonomics and its novelty. We also plan to evaluate this. However, the primary motivation of using a digital pen stems from the spatial and temporal precision of the obtained stroke data which provides the basis for an unprecedented degree of precision during analysing this data for small and subtle patterns; classifying the strokes for their meaning is a sketch interpretation task in addition. As a result, we can get assessment data based on what in written or sketched, and how the spatio-temporal pattern looks like.

## Background and Technology

This research is situated within a long-term project (Sonntag 2015) with the ultimate goal of developing cognitive assistance for patients at home with automatic assessment, compensation and/or training, or assurance/monitoring (figure 1) in terms of cognitive computing.

In addition to understanding people, their processes, their needs, their contexts, in order to create scenarios in which Artificial Intelligence (AI) technology can be integrated, we are particularly concerned to assess and predict the healthcare status with unintrusive sensors (Gelissen and Sonntag 2015). Based on using digital pens in breast imaging for instant knowledge acquisition (Sonntag et al. 2014), where the doctor uses the pen, we now use the digital pen for the patient (Prange et al. 2015) as well.

The goal of the Interakt (interactive cognitive assessment tool) project is to improve the diagnostic process of dementia and other forms of cognitive impairments by digitising and digitalising standardised cognitive assessments. We aim at everyday procedures in hospitals (day clinics) and at home. We base the assessments on the CEARAD test battery (Wolfsgruber et al. 2014) for memory performance and attention; the test is paper-pencil based. We transfer excerpts into the digital world by hand-writing recognition, sketch recognition, and additional new parameters provided by the

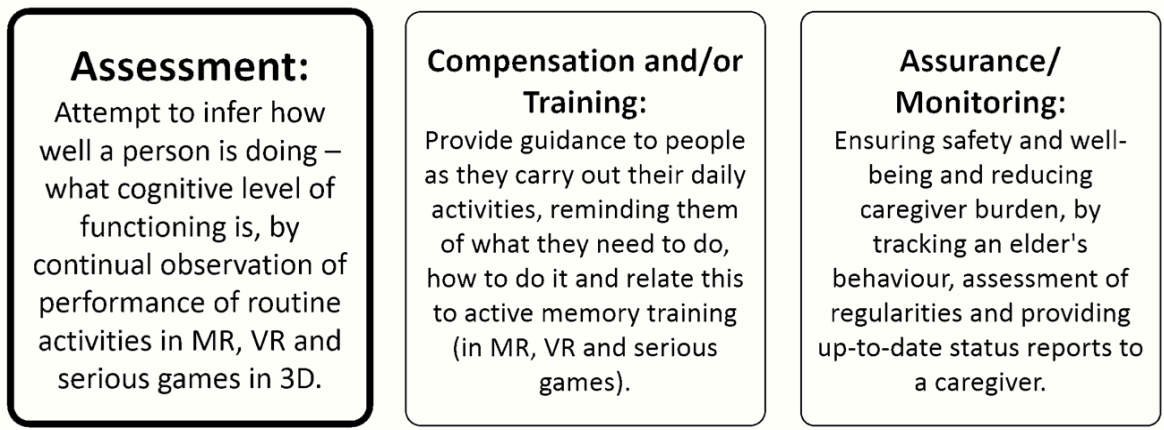

Figure 1: Long-term perspective

digital pen's internal sensors. Using a digital pen has the following potential benefits:

- the caregiver's time to spend on conducting the test can be reduced by automatic assessments;
- the caregiver's attention can be shifted from test features while writing (e.g., easy-to-assess completion of input fields) to verbal test features. The main direction is multimodality (Oviatt et al. 2017);
- the caregiver's time to spend on evaluating the written form can be reduced by automatic assessments of the cognitive state;
- automatic assessments are potentially more objective than human assessments and can include non-stardardised tests and features (for example timing information) whereby previous approaches leave room for different subjective interpretations;
- automatic assessments can use new features of the pen-based sensor environment, to detect and measure new phenomena by more precise measurement. In this way, we need no reliance on a specific clinician's subjective judgment of broad qualitative properties which is also difficult to standardise over longer time periods and multiple tests for a specific patient);
- automatic assessments can, in the future, be conducted in non-clinical environments and at home;
- automatic assessments are relevant for new follow-up checks, they can be conducted and compared in a rigorous and calibrated way;
- automatic assessments can automatically adapt to intrinsic factors (e.g., sensorimotor deficits) if the user model is taken into account;
- automatic assessments reduce extrinsic factors (e.g., misinterpreted verbal instructions);
- automatic assessments allow for evidence of impairment evident in the drawing process (e.g., corrections) instead of static drawings that look normal on paper.

**Research Questions**

The hypothesis is that high-precision data provided by the pen can produce insights about novel metrics, particularly those involving time and those that do no focus solely on properties of the final sketch. A synopsis of the proposed project, including the rationale for the proposed research, and a statement of specific aims and objectives may show the potential significance of the research; a concise point-by-point summary of the aims of the research proposed is enumerated:

1. To identify interface design principles that most effectively support automatic digital assessments by multisensory monitoring and automatic multimodal feedback.

2. At the computational level, it will be important to investigate approaches to designing systems that recognise and model patients' cognitive and affective states with multimodal multisensor technology.

3. At the interface level, it will be important to devise design principles that can inform the development of innovative multimodal-multisensor interfaces for a variety of patient populations, test contexts, and learning environments.

Prospective integrations in the clinical environment as a valid instrument for cognitive assessments require a positive summative evaluation. Preliminary experiments that justify the rationale and demonstrate the feasibility of the proposed research are currently designed. The experimental design and procedures includes the following questions: (1) Which additional knowledge about depression, language capacity, or fine motor skills can be extracted from new multimodal features? (2) How can personalised meta information from written speech or context model data be used to allow for cognitive assessments with finer granularity? (3) How can applied interface design principles of human-computer interaction (GUIs and multimodal interaction) facilitate online and offline interface usability for patients or therapists?

The digital pen analysis assumes content features (what is written, language use, perseveration, i.e., the repetition of a particular response (such as a word, phrase, or gesture) in usual contexts, as well as para-linguistic features (how is it written, style of writing, pauses, corrections, etc.). These are potential technical difficulties and/or limitations in the interpretation of the results.

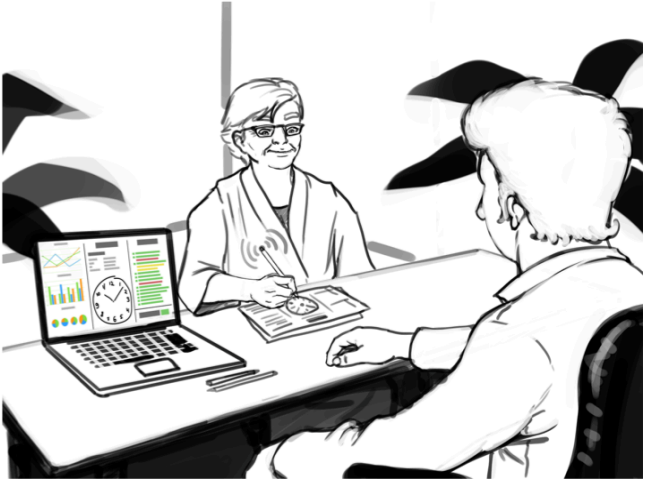

Figure 2: Assessment environment with doctor (right) and patient (left)

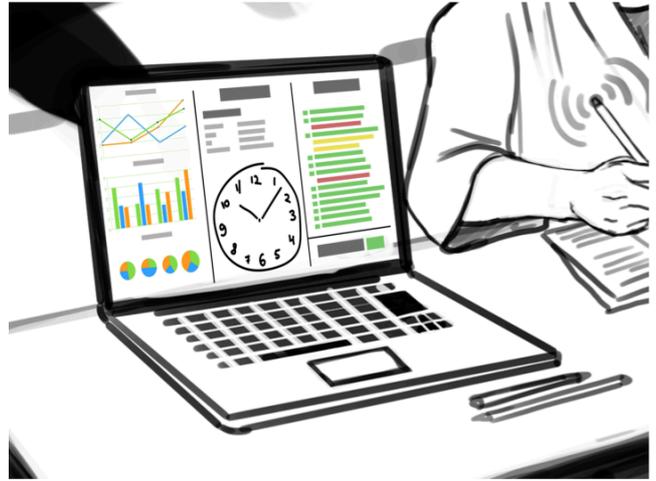

Figure 3: Realtime intelligent user interface for the doctor

Context modelling subsumes the physical activity context and the task context (Bettini et al. 2010). Context attributes are, e.g., context objects, location, and user activity. Therefore, activity recognition has a crucial role to play in our scenarios. AI-based models of context information include software architecture for observing and modelling human activity, to achieve context awareness (Crowley et al. 2002) and additional parameters for clinical assessment.

## Scenarios

The scenario includes the doctor and the patient at a table in a day clinic (figure 2). We implement a sensor network architecture to observe "states" of the physical world and provide real-time access to the data for interpretation. In addition, this context-aware application may need access to a timeline of past events (and world states) in terms of context histories for reasoning purposes while classifying the input data. The result of the real-time assessment of the input stroke data and context data is presented to the doctor in real-time, see figure 3. This description of how we would go about collecting data and test the questions we are examining raise some issues: How will the data from the experiment be gathered? Will it be complete? Will it interfere with the anticipated normal use?

### Technical Architecture

The technical architecture is shown in figure 4. As can be seen, there are two intelligent user interfaces, one for the patient (digital pen interaction) and one for the therapist (caregiver interface). The raw pen data is sent to the document processing and indexing server, and the pen data processing server which provides aggregated pen events in terms of content-based interpretations in RDF. The RDF documents are sent to the data warehouse, together with the RDF meta information. This meta information contains the recognised shapes and text, and text labels, for example. The system attempts to classify each pen stroke and stroke group in a drawing all the time. The second user interface is based on the data warehouse data, and is designed for the practicing clinician; the therapist interface, where the real-time interpretations of the stroke data are available in RDF, is meant to advance existing neuropsychological testing technology. First, it provides captured data in real-time (e.g., for a slow-motion playback), and second, it classifies the analysed high-precision information about both outcome and filling process, opening up the possibility of detecting and visualising subtle cognitive impairment; also it is zoomable to permit extremely detailed visual examination of the data if needed—as previously exemplified in (Davis et al. 2014).

## Future Research

The ultimate goal is to apply the pen-based assessments to treat patients in a semi-automatic (as shown here) and automatic fashion. For interpreting verbal utterances of the CERAD test battery (therapists have problems in taking notes of user answers and comments while conducting a test), a dialogue framework can be used in the future (Neßelrath 2016). Combining speech and pen input (active input) should, in the future, be combined with exploring multimodal approaches to determining cognitive status through the detection and analysis of subtle behaviours and skin conductance sensors (passive input).

## Acknowledgements

This research has been funded by the Federal Ministry of Education and Research (BMBF) under grant number 16SV7768. See the DFKI page http://www.dfki.de/MedicalCPS/?page_id=725 for more information.

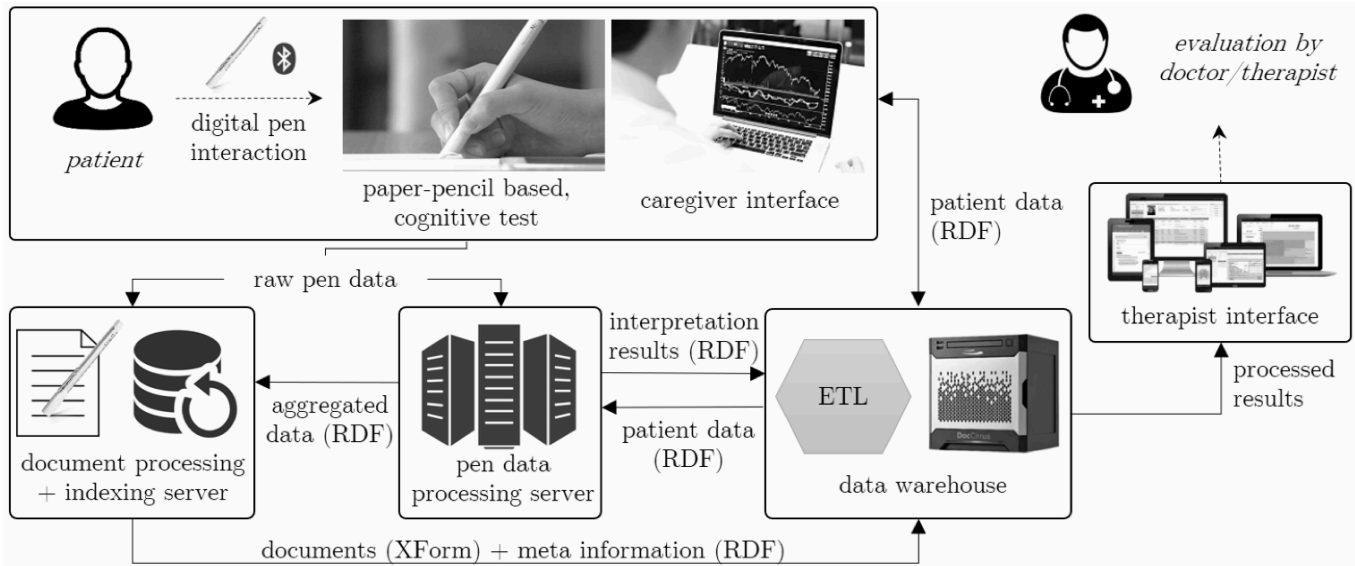

Figure 4: Architecture